\documentclass[aps,pra,showpacs,twocolumn,amsmath,amssymb]{revtex4}
\usepackage{graphicx}
\usepackage{epsfig}
\usepackage{color}
\usepackage{bm}
\usepackage{amssymb}
\usepackage{amsmath}
\usepackage{latexsym}
\usepackage{longtable}
\definecolor{highlight}{rgb}{1.0,0.0,0.0}
\newcommand{\ind}[1]{_{#1}}    
\newcommand{\indrm}[1]{_{\mathrm {#1}}}    
\newcommand{\thick}{d}   
\newcommand{\vc}[1]{\mbox{\boldmath $#1$}} 
\newcommand{\ekspon}[1]{{\mathrm e}^{\textstyle #1}} 

\newcommand{\extleng}{\bar{\Lambda}_{\ind{H}}}
\newcommand{\extlengs}{{\extleng ^{\ind{{\mathrm {\!(s)}}}}}}

\newcommand{\timex}{\mathcal{T}_{\ind{\Lambda}}}
\newcommand{\timexo}{\mathcal{T}_{\ind{0}}}
\newcommand{\delaye}{\tau_{\ind{e}}}

\newcommand{\gammah}{\gamma_{\ind{H}}}   
\newcommand{\gammao}{\gamma_{\ind{0}}}   

\newcommand{\Deltaeh}{\Delta E_{\ind{H}}}
\newcommand{\Deltaeo}{\Delta E_{\ind{0}}}

\newcommand{\refl}{R}

\newcommand{\fieldout}{{\mathcal{E}}}

\newcommand{\incamp}{{\cal E}_{\indrm{SASE}}}

\newcommand{\thicknessduration}{\mathcal{T}_{\ind{d}}}
\newcommand{\sgn}[1]{{\mathrm {sgn}}\left\{ {#1} \right\}}

\newcommand{\spflux}{F}
\newcommand{\spfluxnorm}{\bar{F}}
\newcommand{\xduration}{\Delta t_{\indrm{x}}}
\newcommand{\xband}{\Delta E_{\indrm{x}}}
\newcommand{\xdurationrms}{\sigma_{\indrm{x}}}
\newcommand{\xbandrms}{\sigma_{\indrm{E_{\mathrm x}}}}
\newcommand{\bunchlengthrms}{\sigma_{\indrm{e}}}
\newcommand{\bunchduration}{\Delta t_{\indrm{e}}}
\begin{document}
\title{Maximizing Spectral Flux from Self-Seeding Hard X-ray  FELs}
\author{Xi  Yang}\email{xiyang@bnl.gov} 
\affiliation{Brookhaven National Laboratory, Upton, NY 11973, USA}
\author{Yuri Shvyd'ko}\email{shvydko@aps.anl.gov}
\affiliation{Advanced Photon Source, Argonne National Laboratory,
  Argonne, Illinois 60439, USA}

\date{\today}
\begin{abstract} 
  Fully coherent x-rays can be generated by self-seeding x-ray
  free-electron lasers (XFELs).  Self-seeding by a forward Bragg
  diffraction (FBD) monochromator has been recently proposed
  \cite{GKS11} and demonstrated \cite{HXRSS12}. Characteristic time
  $\timexo$ of FBD determines the power, spectral, and time
  characteristics of the FBD seed \cite{SL12}. Here we show that for a
  given electron bunch with duration $\bunchlengthrms $ the spectral
  flux of the self-seeding XFEL can be maximized, and the spectral
  bandwidth can be respectively minimized by choosing $\timexo\sim
  \bunchlengthrms/\pi $ and by optimizing the electron bunch delay
  $\delaye$.  The choices of $\timexo$ and $\delaye$ are not
  unique. In all cases, the maximum value of the spectral flux and the
  minimum bandwidth are primarily determined by $\bunchlengthrms
  $. Two-color seeding takes place if $\timexo\ll\bunchlengthrms/\pi
  $. The studies are performed, for a Gaussian electron bunch
  distribution with the parameters, close to those used in the
  short-bunch ($\bunchlengthrms \simeq 5$~fs) and long-bunch
  ($\bunchlengthrms \simeq 20$~fs) operation modes of the LCLS XFEL.
\end{abstract}
%
%
%
\maketitle

\section{Introduction}
Seeded x-ray free-electron lasers (XFELs) \cite{HXRSS12,FERMI12}
generate fully coherent x-rays with a well-defined spectrum and higher
spectral flux.  Their supreme spectral properties will expand the
science reach of the XFELs, and stimulate the utilization of advanced
high-resolution spectroscopic techniques.  The self-seeding XFEL is
also a path to increase the peak radiation power of XFELs to the TW
levels \cite{JWC12}.  In the present paper we study the conditions for
stable seeding, with the smallest shot-to-shot variations of output
radiation, the narrowest bandwidth, and largest spectral flux.

\begin{figure}[t!]
\setlength{\unitlength}{\textwidth}
\begin{picture}(1,0.22)(0,0)
\put(0.0,0.0){\includegraphics[width=0.5\textwidth]{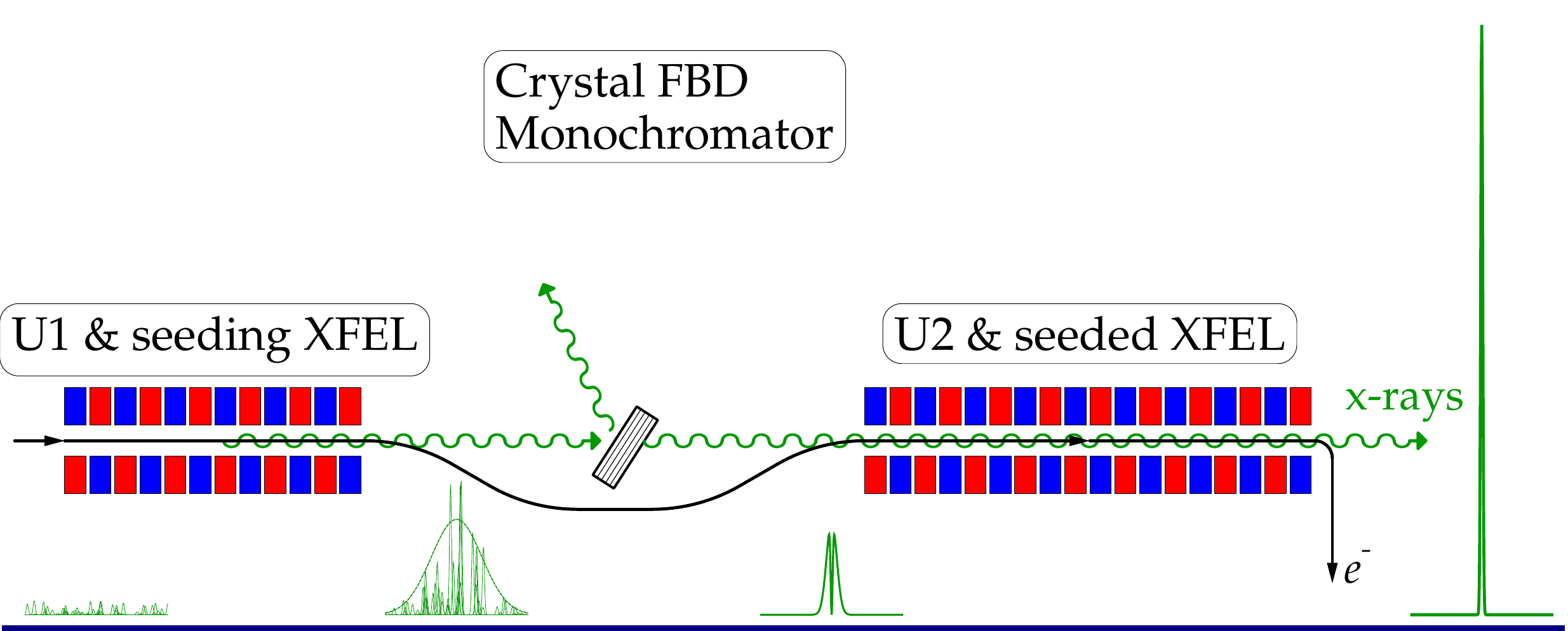}}
\end{picture}
\caption{(Color online) Hard x-ray FEL self-seeding scheme uses x-rays from the first
  half of the magnetic undulator system U1 to seed the electron bunch
  in the second half U2 via a single crystal x-ray monochromator. The
  monochromator produces the delayed monochromatic seed under the
  forward Bragg diffraction (FBD) conditions \cite{GKS11}.}
\label{Fig001}
\end{figure}

The self-seeding scheme uses x-rays from the first half of the FEL
system (electron beam -- magnetic undulator system -- radiation field)
to generate radiation by the self-amplified spontaneous emission
(SASE) process \cite{DKS82,BPN84,Kim86,SSY,HK07} to seed the electron
bunch in the second half of the FEL system via an x-ray monochromator
\cite{FSS97,SSSY}.  The delay of x-rays in the monochromator is
compensated by the appropriate delay $\delaye$ of the electron bunch
in the magnetic chicane. Standard x-ray monochromators produce delays
in the picosecond range, at least.  In the hard x-ray regime with
high-energy electrons, a large magnetic chicane is required to match
the path-length delay of the x-ray monochromator, which may degrade
the electron-beam qualities.  A possible solution is using a
two-electron-bunch self-seeding scheme \cite{DZR10}. Another elegant
solution, proposed by Geloni, Kocharyan, and Saldin, uses forward
Bragg diffraction (FBD) of x-rays from a diamond crystal to generate a
monochromatic seed \cite{GKS11} - Fig.~\ref{Fig001}.  The first
trailing maximum of FBD was proposed for the self-seeding of $\simeq
5$-fs (FWHM) short electron bunches, at an optimal $\approx 20$-fs
delay of the seed relative to the electron bunch.  The proposal have
been recently realized and fully coherent hard x-rays have been
generated at the LCLS XFEL by an international team led by Emma
\cite{HXRSS12}.

The question that naturally arises is: Under what conditions is
seeding most efficient in terms of achieving the shortest XFEL
saturation length, the smallest spectral bandwidth of output
radiation, and the largest spectral density?  One could expect that
for this, generally speaking, the monochromatic seed duration should
be comparable to the duration of the electron pulse, the monochromatic
seed power should substantially exceed the shot noise power, and the
peak seed power should overlap with the peak electron current density.
The question then concretizes to which crystal, which Bragg
reflection, which scattering geometry, which crystal thickness, etc.
suit best? This is a problem with many parameters, which is difficult
to tackle. However, there is a possibility to reduce the problem to
one single parameter.

\begin{figure*}[t!]
\setlength{\unitlength}{\textwidth}
\begin{picture}(1,0.22)(0,0)
\put(0.00,-0.02){\includegraphics[width=1.0\textwidth]{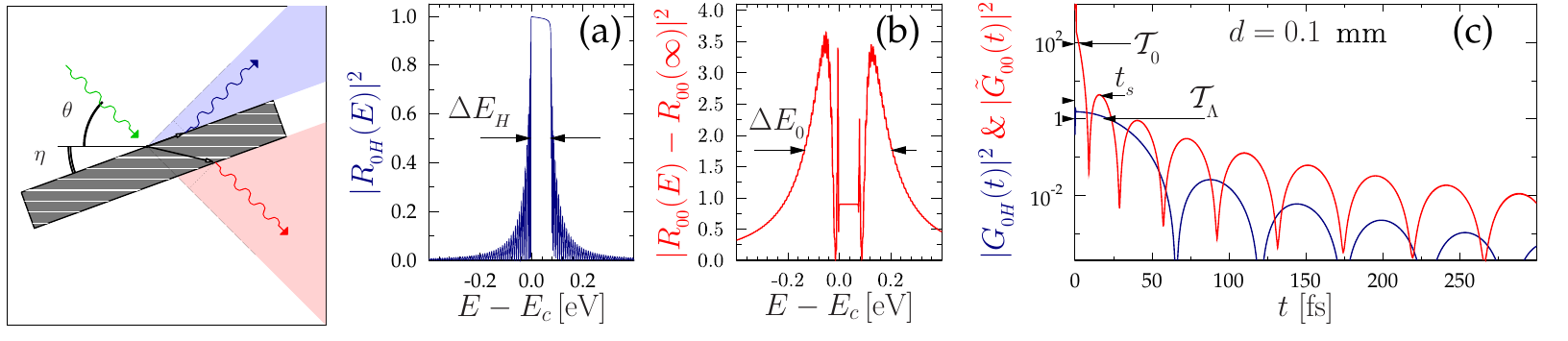}}
\end{picture}
\caption{(Color online) (left) Schematic of Bragg diffraction (BD) and forward Bragg
  diffraction (FBD) of x-rays from a crystal in Bragg-case geometry.
  (a) Example of energy dependence of x-ray reflectivity
  $|\refl_{\ind{0H}}(E)|^2$ in Bragg diffraction and (b) of energy
  dependence of forward Bragg diffraction
  $|\refl_{\ind{00}}(E)-\refl_{\ind{00}}(\infty)|^2$ from a diamond
  crystal in the 004 Bragg reflection, $E_{\indrm{c}}=8.33$~keV,
  $\theta=56.6^{\circ}$, $\eta=0^{\circ}$, $\thick=0.1$~mm. (c)
  Intensities of corresponding time dependencies of crystal response
  to an excitation with ultra-short x-ray pulse in BD
  $|G_{\ind{0H}}(t)|^2$ (blue) and in FBD $|G_{\ind{00}}(t)|^2$ (red),
  respectively.}
\label{Fig002}
\end{figure*}

It has been shown in \cite{SL12} that under a certain approximation
the time dependence of FBD can be characterized by a single parameter
$\timexo$, the time constant of forward Bragg diffraction.  It is the
main parameter that defines the strength, and duration of the FBD
monochromatic seed.  It is a generic feature for all symmetric or
asymmetric, for all transmission (Laue-case) or reflection
(Bragg-case) scattering geometries. The physics is controlled by the
parameters that compose $\timexo$: the magnitude of the effective
crystal thickness (crystal thickness seen by the incident beam), and
the extinction length in the symmetric Bragg reflection
(characteristic length of Bragg diffraction). This approximation is
correct if we are interested in the FBD radiation emanating from the
crystal with delays $t$, which are significantly shorter than a mere
propagation time $\thicknessduration$ of the radiation through the
crystal. It turns out that exactly for self-seeding applications the
approximation $t \ll \thicknessduration $ holds well in many cases.

Due to this property, the spectral flux optimization could be studied
by analyzing self-seeding as a function of the principle time constant
$\timexo$. With one more time constant $\thicknessduration$ taken into
account, a more accurate picture can be achieved. Here we perform such
studies using the 3D time-dependent FEL simulation code GENESIS 1.3 by
Sven Reiche \cite{GENESIS}.

The paper is organized as follows.  In Sec.~\ref{seed-calculations} we
briefly summarize the basic properties of Bragg diffraction (BD) and
the complementary to it forward Bragg diffraction (FBD), and outline
the procedure to calculate the monochromatic seed in forward Bragg
diffraction from a crystal in the XFEL self-seeding scheme. In
Sec.~\ref{3dsimulations} we provide details on the 3D FEL
simulations. In Sec.~\ref{results} we present and discuss the results
of the simulations.

\section{Generation of the Monochromatic Seed in FBD}
\label{seed-calculations}

\subsection{Basic Properties of Forward Bragg Diffraction} 

Here we briefly summarize the basic properties of Bragg diffraction
(FB) and of the complementary forward Bragg diffraction (FBD); define
their characteristic spectral, time, and space parameters; and
outline, using results and notations of \cite{SL12}, the procedure for
calculating the monochromatic seed in FBD from a crystal in the XFEL
self-seeding scheme.

Bragg diffraction (BD) of x-rays incident upon a crystal at a glancing
angle $\theta$ to reflecting atomic planes with diffraction vector
$\vc{H}$, see Fig.~\ref{Fig002} (left), takes place within a spectral
range $\Deltaeh$ centered at a photon energy
$E_{\indrm{c}}=Hc\hbar/2\sin\theta$, defined by Bragg's law.  Here $c$
is the speed of light in vacuum, and $\hbar$ is Planck's constant.  An
example Bragg reflection spectral dependence $|\refl_{\ind{0H}}(E)|^2$
is shown in Fig.~\ref{Fig002}(a).  All photons within the Bragg
reflection spectral range $\Deltaeh$ are almost totally reflected from
the crystal and propagate at a scattering angle $2\theta$ with a
characteristic delay time $\timex=2\hbar/\Deltaeh$ of Bragg
diffraction, as shown in Fig.~\ref{Fig002}(c). In fact, $\timex$
scales with the characteristic length of Bragg diffraction $\extleng$,
the so-called extinction length:
\begin{equation}
  \timex\,=\,
\frac{2\,\extleng\, \sin^2\theta}{c\, |\gammah| }, \hspace{1cm} \timex\, \Deltaeh=2\hbar .
\label{timex}
\end{equation}
Here $\gammah=\sin(\eta-\theta)$, and $\eta$ is an asymmetry angle,
see Fig.~\ref{Fig002} (left).

\begin{figure*}[t!]
\setlength{\unitlength}{\textwidth}
\begin{picture}(1,0.20)(0,0)
\put(0.00,-0.02){\includegraphics[width=1.0\textwidth]{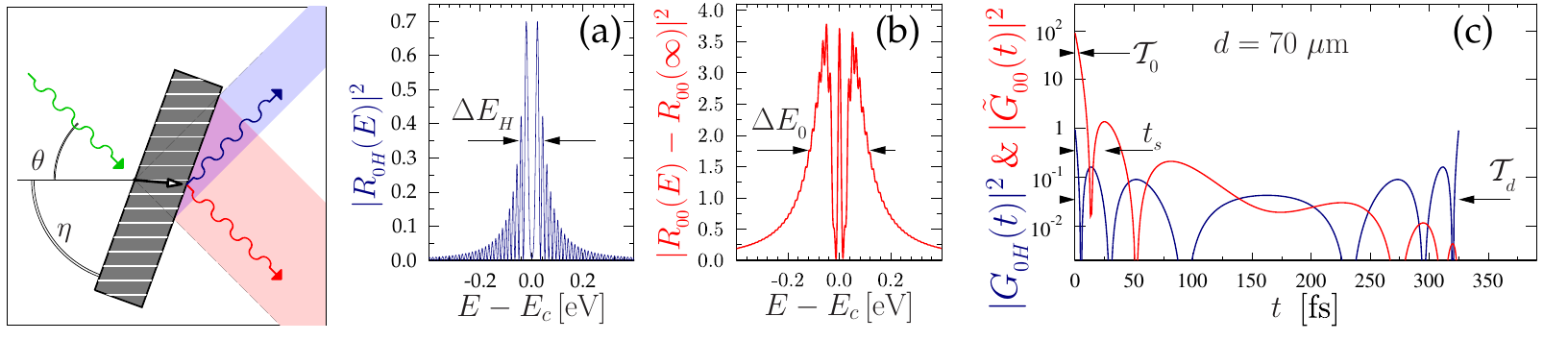}}
\end{picture}
\caption{(Color online) (left) Schematic of Bragg diffraction (BD) and forward Bragg
  diffraction (FBD) of x-rays from a crystal in Laue-case scattering
  geometry.  All notations are as in Fig.~\ref{Fig002}.  Example of
  energy and time dependencies from a diamond crystal in the 004 Bragg
  reflection, $E_{\indrm{c}}=8.33$~keV, $\theta=56.6^{\circ}$,
  $\eta=90^{\circ}$ $\thick=0.1$~mm. }
\label{Fig003}
\end{figure*}

Forward Bragg diffraction (FBD) is a multiple coherent scattering of
x-rays from the same atomic planes. It is complementary to the BD
scattering process. Those spectral components of the incident
radiation contribute to FBD, which do not participate in BD, those
which are outside but in the immediate vicinity of the Bragg
reflection region $\Deltaeh$. The FBD spectral dependence
$|\refl_{\ind{00}}(E)-\refl_{\ind{00}}(\infty)|^2$ in
Fig.~\ref{Fig002}(b) illustrates this. Here $\refl_{\ind{00}}(E)$ is a
transmission amplitude through the crystal of a spectral component
with energy $E$. The FBD spectral width $\Deltaeo$
scales with the BD spectral width $\Deltaeh$,
is larger than $\Deltaeh$, and increases with the crystal thickness
$\thick$: 
\begin{equation}
\Deltaeo = \Deltaeh\frac{\thick}{2\pi\extleng}, \hspace{0.5cm} 
\timexo={\timex}\frac{\extleng}{\thick}, \hspace{0.5cm} \timexo\, \Deltaeo=\frac{\hbar}{\pi}.
\label{deltaeo-timexo}
\end{equation}
Respectively, the characteristic time of FBD $\timexo$
see Fig.~\ref{Fig002}(c), scales
with $\timex$, is smaller than $\timex$, and decreases with the crystal
thickness.

The FBD crystal response to the excitation by an incident radiation
pulse can be calculated as a convolution in time of the incident SASE
radiation pulse amplitude
$\incamp(t)$ 
and of a forward Bragg diffraction response function $G_{\ind{00}}$:
\begin{equation}
\fieldout_{\ind{0}}(t)\,=\, 
\int_{-\infty}^{t}{{\mathrm d}t^{\prime}}\, 
G_{\ind{00}}(t-t^{\prime}) \incamp (t^{\prime}). \label{pro026}
\end{equation}
Here $\incamp(t)$ is a slowly varying envelope.
$G_{\ind{00}}(t)$ presents the crystal
response to a delta-function excitation  in time and is calculated as
a Fourier transform of the FBD spectral amplitudes:
\begin{equation}\label{rta092}
\begin{split}
G_{\ind{00}}(t)  \,= & \, \refl_{\ind{00}}(\infty)\, \left[ \delta(t)\,+\, \tilde{G}_{\ind{00}}(t)\right]\\
\tilde{G}_{\ind{00}}(t)\,= & \,\int_{-\infty}^{\infty}\frac{{\mathrm
      d}\Omega}{2\pi}\,\ekspon{-{\mathrm i}\Omega t}\,
\frac{\refl_{\ind{00}}(\Omega)-\refl_{\ind{00}}(\infty)}{\refl_{\ind{00}}(\infty)},
\end{split}
\end{equation}
where $\Omega= (E-E_{\indrm{c}})/\hbar$.  According to
Eq.~\eqref{rta092}, $G_{\ind{00}}(t)$ is in fact a sum of the prompt,
i.e., diffraction-free transmission $\refl_{\ind{00}}(\infty)
\delta(t)$ of the far-from-Bragg-reflection-region spectral
components, and of the delayed {\em actual} forward diffraction
response function $\tilde{G}_{\ind{00}}(t)$.  As a result, the final
expression for the time dependence of the radiation field emanating
from the crystal in the forward direction
\begin{equation}
\fieldout_{\ind{0}}(t)=
\refl_{\ind{00}}(\infty)\left[\incamp(t)+\!
 \int_{-\infty}^{t}\!\!\!{{\mathrm d}t^{\prime}} 
\tilde{G}_{\ind{00}}(t-t^{\prime}) \incamp (t^{\prime})\right]
\label{pro076}
\end{equation}
is a sum of the diffraction-free transmitted SASE radiation and of
the second component, which is the delayed monochromatic seed due to
FBD.

In general, FBD is characterized also by a spatial transverse shift of
the delayed radiation emitted by the crystal in the forward direction
\cite{LS12,SL12}.  In the present studies we will, however, assume for
simplicity that the spatial shift of the FBD signal is small and can
be neglected.

\begin{figure}[t!]
\setlength{\unitlength}{\textwidth}
\begin{picture}(1,0.34)(0,0)
\put(0.0,0.00){\includegraphics[width=0.5\textwidth]{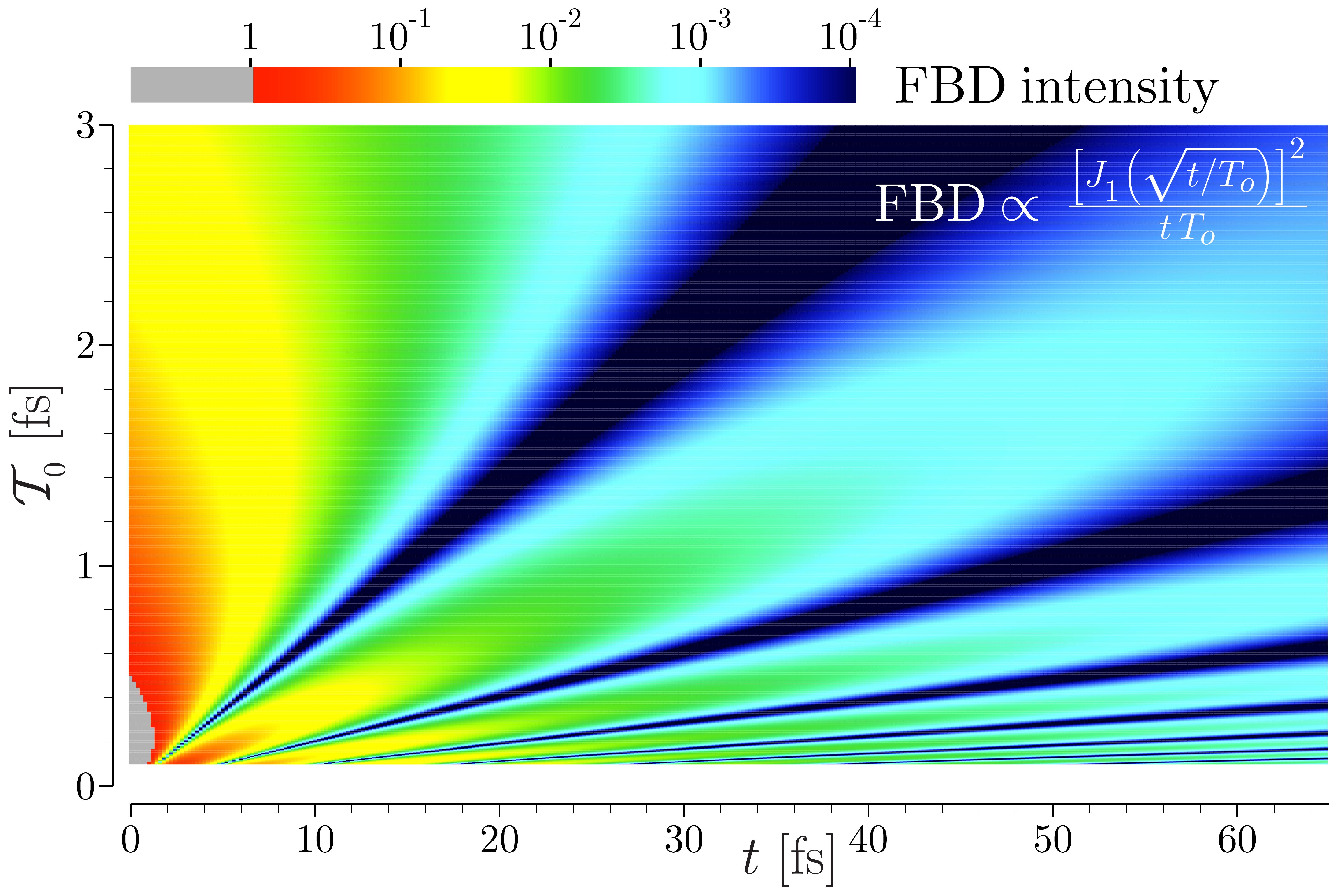}}
\end{picture}
\caption{(Color online) 2D plot of the universal FBD response function intensity
  $|\tilde{G}_{\ind{00}}(t)|^2$ (in relative units) plotted for
  different values of the time delay $t$ to the instantaneous
  excitation of the crystal, and for different values of the FBD time
  constant $\timexo$, using Eq.~\eqref{rta1250}. }
\label{Fig004}
\end{figure}

An analytical expression can be derived for the forward diffraction
time response function $\tilde{G}_{\ind{00}}(t)$ in the approximation
of a non-absorbing and thick $\thick \gg \extlengs$ crystal valid in
the general case of asymmetric Bragg diffraction, both in Bragg-case
geometry (Fig.~\ref{Fig002}):
\begin{equation}\label{rta125}
  \tilde{G}_{\ind{00}}(t) 
=  -\frac{1}{2\timexo}\,
\frac{J_{\ind{1}}\left[\sqrt{\frac{t}{\timexo} \left(1+\frac{t}{\thicknessduration}\right)}\, \right]}
{\sqrt{\frac{t}{\timexo}\left(1+\frac{t}{\thicknessduration}\right)}},
\end{equation}
and in Laue-case geometry (Fig.~\ref{Fig003}):
\begin{equation}\label{lta050}
\begin{split}
\tilde{G}_{\ind{00}}(t)=&\frac{1}{2\timexo}  \left(1-\frac{t}{\thicknessduration} \right) 
\frac{J_{\ind{1}}\left[\sqrt{\frac{t}{\timexo} \left(1-\frac{t}{\thicknessduration}\right)} \right]}{\sqrt{\frac{t}{\timexo} \left(1-\frac{t}{\thicknessduration}\right)}} 
\\
&[0< t  < \thicknessduration ],   
\end{split}
\end{equation}
with the characteristic time parameters 
\begin{equation}\label{timexo}  
  \timexo=
\frac{2[\extlengs]^2}{c (\thick/\gammao)}, \hspace{0.5cm}
\thicknessduration  =  \frac{2\,\thick\,\sin^2\theta}{c|\gammah |}.\\
\end{equation}
$\timexo$ is the characteristic time of forward Bragg diffraction, the
same as in Eq.~\eqref{deltaeo-timexo}. Here it is expressed, however,
through a Bragg-reflection-invariant -- the extinction length
$\extlengs$ in symmetric diffraction geometry ($\eta=0$) --  and by the
effective crystal thickness $\thick/\gammao$, the thickness seen by
incident x-rays, $\gammao=\sin(\theta+\eta)$. Extinction length
$\extlengs$ values for Bragg reflections in diamond, Si, and
Al$_2$O$_3$ are tabulated in \cite{Shvydko-SB,SL12}.  The parameter
$\thicknessduration$ is a characteristic measure of time in Bragg
diffraction associated with the crystal thickness $\thick$. In the
Laue-case geometry, $\thicknessduration$ is equal to the total
duration of forward Bragg diffraction, see  Fig~\ref{Fig004}(c). In the
Bragg-case geometry, it measures the time between multiple
reflections.

If additionally, the delay times of interest are short $t\ll
\thicknessduration$ compared to the propagation time through the
crystal, a generic expression can be used for all symmetric or
asymmetric,
Brag-case or Laue-case geometries: 
\begin{equation}\label{rta1250}
  \tilde{G}_{\ind{00}}(t) 
= \sgn{b} \frac{1}{2\timexo}\,
\frac{J_{\ind{1}}\left(\sqrt{{t}/{\timexo} }\, \right)}
{\sqrt{{t}/{\timexo}}}, \hspace{0.5cm}t \ll \thicknessduration .
\end{equation}
The time constant of forward Bragg diffraction $\timexo$
Eq.~\eqref{timexo} is the only parameter that enters
Eq.~\eqref{rta1250}, and is the only parameter that defines the
strength, delay, and duration of FBD and therefore of the
monochromatic seed.  The physics is controlled by the parameters that
compose $\timexo$ Eq.~\eqref{timexo}: the magnitude of the effective
crystal thickness $\thick/\gammao$, and the extinction length
$\extlengs$ in the symmetric Bragg reflection.  A 2D plot of the FBD
intensity $\propto |\tilde{G}_{\ind{00}}(t)|^2$, in the ($t,\timexo$)
space, shown in Fig.~\ref{Fig004}, can be used as an initial guide to
the most beneficial $\timexo$ values for seeding at a given time delay
$t$.

\begin{table}[b!]
\centering
\begin{tabular}{|l|l|l|llll||ll|l|}
  \hline  \hline  
$\vc{H}$  & $\extlengs$ & $\thick$ & $\theta$ & $\eta$ & $\gammah$ & $\gammao$ & $\timexo$ & $\thicknessduration$ & $\Deltaeo$ \\
$hkl$  & $\mu$m & $\mu$m & deg & deg &  &  & fs  & fs & meV \\
  \hline  \hline  
111  & 1.09 & 40 & 21.3  & 0 & 0.36  & -0.36 & 0.07  & 94 & 2993 \\
111  & 1.09 & 100 & 21.3  & 90 & 0.93  & 0.93 & 0.07  & 94 &  \\
220  & 1.98 & 75 & 36.3 & 0 & 0.59 & -0.59  & 0.21  & 296 & 998 \\
220  & 1.98 & 100 & 36.3 & 90 & 0.81 & 0.81  & 0.21  & 296 & \\
400  & 3.63 & 100 & 56.8 & 0  & 0.84 & -0.84  & 0.73  & 558 & 287 \\
511  & 7.83 & 260 & 82.9 & 15.8 & 0.99 & -0.92 & 1.5  & 1720 & 140 \\
511  & 7.83 & 130 & 82.9 & 15.8 & 0.99 & -0.92 & 3.0  & 860 & 70\\
511  & 7.83 & 100 & 82.9 & 15.8 & 0.99 & -0.92 & 4.1  & 712 & 51 \\
511  & 7.83 & 58 & 82.9 & 15.8 & 0.99 & -0.92 & 7.0  & 413 & 30 \\
 \hline  \hline    
\end{tabular}
\caption{Examples of  the  time constants  $\timexo$, 
  and   $\thicknessduration$ in Eq.~\eqref{timexo} used in the subsequent simulations; and their realization through 
  particular (not at all unique)  Bragg diffraction cases. The appropriate FBD spectral bandwidths\ $\Deltaeo$ 
  -- Eq.~\eqref{deltaeo-timexo} -- are also provided for reference. 
  Each  Bragg diffraction case is defined first of all by the 
  Bragg reflection invariant  diffraction vector
  $\vc{H}$ and symmetric extinction length $\extlengs$ in diamond crystals, 
  by the crystal thickness $\thick$, and by the parameters defining scattering geometry: 
  the Bragg angle $\theta$, the asymmetry angle 
  $\eta$, the incidence wave vector projection $\gammao=\sin(\theta+\eta)$   to the crystal internal normal, 
  and the reflection wave vector projection 
  $\gammah=\sin(\eta-\theta)$, respectively. 
}
\label{tab0}
\end{table}

%
%

The first trailing maximum of $|\tilde{G}_{\ind{00}}(t)|^2$, which was
originally proposed to use as the monochromatic seed, appears at the
time delay $t_{\ind{s}}=26\timexo$, see Fig.~\ref{Fig002}(c). Its
duration is $\Delta t_{\ind{s}}=16.5\timexo$. Both the delay and
duration can be tailored by changing $\timexo$, which can be done
practically by adjusting the extinction length $\extlengs$ (for
example, by choosing another reflection or asymmetry parameter), or by
changing the crystal thickness.

Table~\ref{tab0} shows examples of $\timexo$ and $\thicknessduration$
used in the subsequent simulations and how these values can be
realized with particular Bragg reflections. The realization of a
particular $\timexo$ is not unique. The same $\timexo$ values can be
realized with different Bragg diffraction vectors, i.e., different
$\extlengs$, different crystal thicknesses $\thick$, and different
scattering geometry parameters $\theta$, $\eta$, $\gammah$, and $\gammao$.

\begin{figure}[t!]
\setlength{\unitlength}{\textwidth}
\begin{picture}(1,0.34)(0,0)
\put(0.0,0.00){\includegraphics[width=0.5\textwidth]{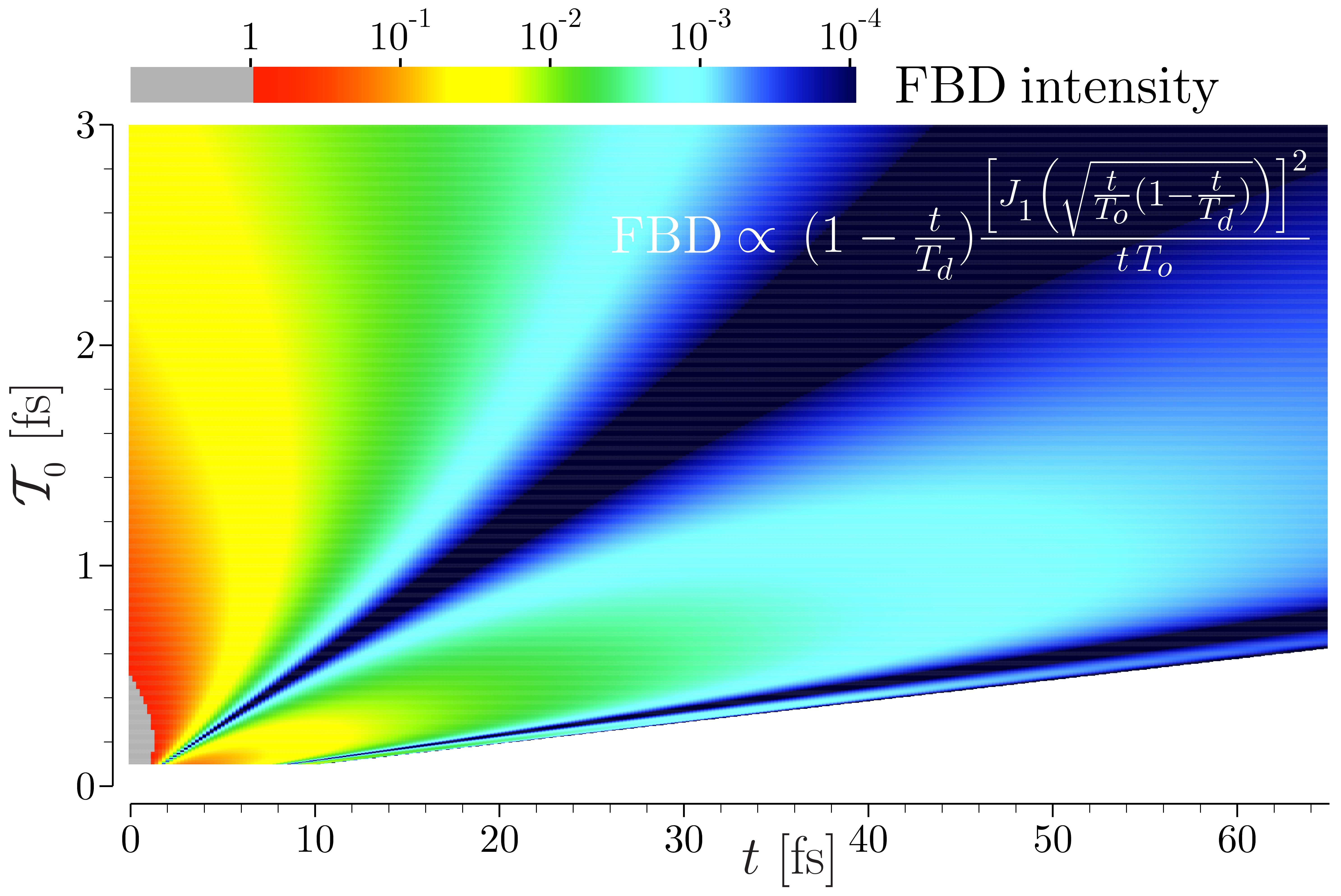}}
\end{picture}
\caption{(Color online) Similar to Fig.~\eqref{Fig004}, however, Eq.~\ref{lta050} for
  the Laue-case FBD response function is used for a particular case of
  $\thicknessduration\simeq 104 \timexo$, which permits the duration
  of the FBD peaks to be stretched. }
\label{Fig005}
\end{figure}

An interesting situation may occur in Laue-case diffraction. Unlike
the Brag-case diffraction presented by Eq.~\eqref{rta125}, the
argument $a={t}/{\timexo} \left(1-{t}/{\thicknessduration}\right)$ of
the response function in the Laue-case diffraction,
Eq.~\eqref{lta050}, attains the maximum value
$a_{\indrm{max}}={\thicknessduration}/{4\timexo}$ at
$t_{\indrm{max}}=\thicknessduration/2$. The argument changes slowly
with time around $t_{\indrm{max}}$. If in addition
$a_{\indrm{max}}\simeq 26$, i.e., $\thicknessduration/\timexo= (\thick/\extlengs)^2\,\sin^2\theta /(|\gammah|\gammao ) \simeq
104\timexo$, the first trailing maximum of the response function will
be stretched in time compared to the Bragg-case FBD, as graphically
shown in Fig.~\ref{Fig005}. This property maybe useful for seeding
long electron bunches.

\
\begin{figure*}[t!]
\setlength{\unitlength}{\textwidth}
\begin{picture}(1,0.92)(0,0)
\put(0.0,0.00){\includegraphics[width=1.0\textwidth]{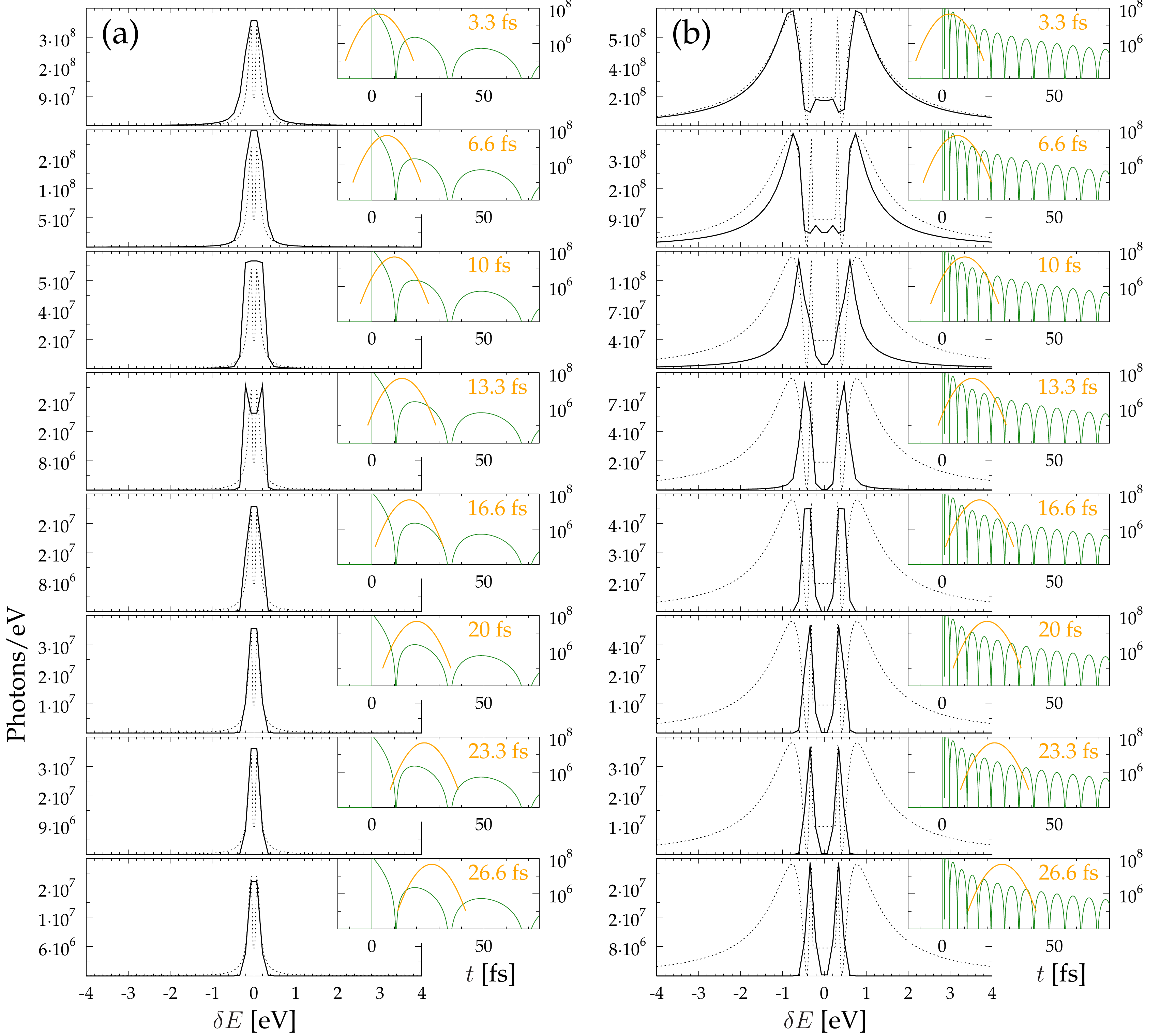}}
\end{picture}
\caption{(Color online) Spectra of the seeding radiation as ``seen'' by the electron
  bunch (solid black lines). The spectra are calculated for different
  delays $\delaye$ of the electron bunch (solid orange lines in the
  insets) with respect to the FBD intensity (solid green lines in the
  insets). The dotted lines show for reference the time-integrated FBD
  spectra. Calculations are presented for an electron bunch with an
  rms duration $\bunchlengthrms=5.3$~fs, Bragg-case FBD time constant
  $\timexo=0.75$~fs (a), and $\timexo=0.07$~fs (b), respectively. Case
  (a) with $ \timexo \simeq \bunchlengthrms/\pi$ is optimal for the
  single-color seeding, while case (b) with $ \timexo \ll
  \bunchlengthrms/\pi$ is better suited for the two-color seeding.}
\label{Fig006}
\end{figure*}

\subsection{Single-Color and Multi-Color Seeding Conditions} 
\label{optimal}

The FBD spectra presented in Figs.~\ref{Fig002}(b) and \ref{Fig003}(b)
are time integrated.  The actual spectrum of the radiation that seeds
the electron bunch is related to it, but not identical, as the
electron bunch ``sees'' only a part of the FBD signal.  The seeding
radiation spectrum is determined by the Fourier transform of the
product of the FBD amplitude $\fieldout_{\ind{0}}(t)$ -
Eq.~\eqref{pro076}, and of the electron current time profile.
Examples of seeding radiation spectra, calculated for an idealistic
case of prompt excitation $\incamp(t)\propto \delta(t)$ and for an
electron bunch with Gaussian profile are shown in Fig.~\ref{Fig006}.
The radiation spectrum that seeds the electron bunch depends largely
on the relationship between the electron bunch duration
$\bunchlengthrms$ and the time profile of the radiation
overlapping with the electron bunch. The latter can be varied either
by the FBD time constant $\timexo$ or/and by the electron bunch delay
$\delaye$.

In the limiting case of a flat time profile, the seeding spectrum is
defined by $\bunchlengthrms$ and will have a spectral width of $\simeq
\hbar/ \bunchlengthrms$.  This spectral width represents the finest
spectral feature that the XFEL system can generate. This is also a
typical spectral width of the spikes in the SASE spectra
\cite{SSY98,HK07}.

Therefore, if the FDB spectral width is chosen to be $\Deltaeo \simeq \hbar/
\bunchlengthrms $, i.e., $\bunchlengthrms \simeq \pi \timexo $, as in
the case presented in Fig.~\ref{Fig006}(a), a close to optimal
condition for single-color seeding can be achieved.

In the other limiting case, $\Deltaeo \gg \hbar/ \bunchlengthrms $,
i.e., $\bunchlengthrms \gg \pi \timexo $, the FBD time profile has a
few oscillations superimposed on the electron bunch. As a result, the
seeding spectrum will have a double-peak structure with the separation
between the peaks inversely proportional to the oscillation period,
i.e., to $\propto 1/\timexo$.  This case is presented in
Fig.~\ref{Fig006}(b). The FBD spectrum seen by the electron bunch
changes dramatically with time. At small time delays the two outer
broader peaks of the FBD spectrum dominate. At larger time delays, the
two narrow spectral peaks survive and represent the seeding spectrum.
In any case, at any time, the double-peak structure persists, a
prerequisite for the two-color seeding. For this, however, appropriate
spectral components should also be present in the SASE spectrum.  Due
to the stochastic nature of the SASE spectrum, this likely will not
always be the case. The XFEL output will result in a two-color output,
however, with the spectral content varying from shot to shot.

The Laue-case FBD spectrum can be more fine-structured, see
Fig.~\ref{Fig003}(b). This may provide conditions for more than
two-color seeding, i.e., multi-color seeding.  Even more important, in
the Laue-case geometry the fine structure may have a single narrow
peak in the center (unlike the Bragg-case, which always has a
double-hump structure); this could be beneficial for a true
single-color seeding, as the broad-band spectral components decay
fast, leaving only the sharp spectral component for single-color
seeding.

\section{3D FEL Simulation Details}
\label{3dsimulations}

The 3D time-dependent FEL simulation code GENESIS 1.3 \cite{GENESIS}
is applied to study XFEL self-seeding with the FBD (wake)
monochromator. The main FEL parameters used in the simulations are
presented in Table~\ref{tab1}. They are based on the LCLS
commissioning results \cite{EAA10}. Two magnetic undulator systems are
used, both with the nominal undulator period
$\lambda_{\ind{w}}=30$~mm. The length of the first undulator system U1
is chosen to ensure FEL operations in the exponential growth regime
and to achieve the x-ray seed power after the monochromator about two
orders of magnitude larger than the shot noise power $P_{\indrm{n}}$.
At the same time, the U1 length is limited to 60~m to avoid energy
spread growth due to FEL saturation. The second undulator system U2 is
chosen to be 70~m long for the FEL to reach saturation with reduced
fluctuations of the x-ray pulse power. The lengths of the first and
the second magnetic undulator systems include the gaps between the
undulator sections.  A weak four-dipole chicane with $R_{56} \simeq
3-45~\mu$m is installed in a 4-m-long space between the U1 and U2
undulator systems.

\begin{figure}[t!]
\setlength{\unitlength}{\textwidth}
\begin{picture}(1,0.35)(0,0)
\put(0.0,0.00){\includegraphics[width=0.5\textwidth]{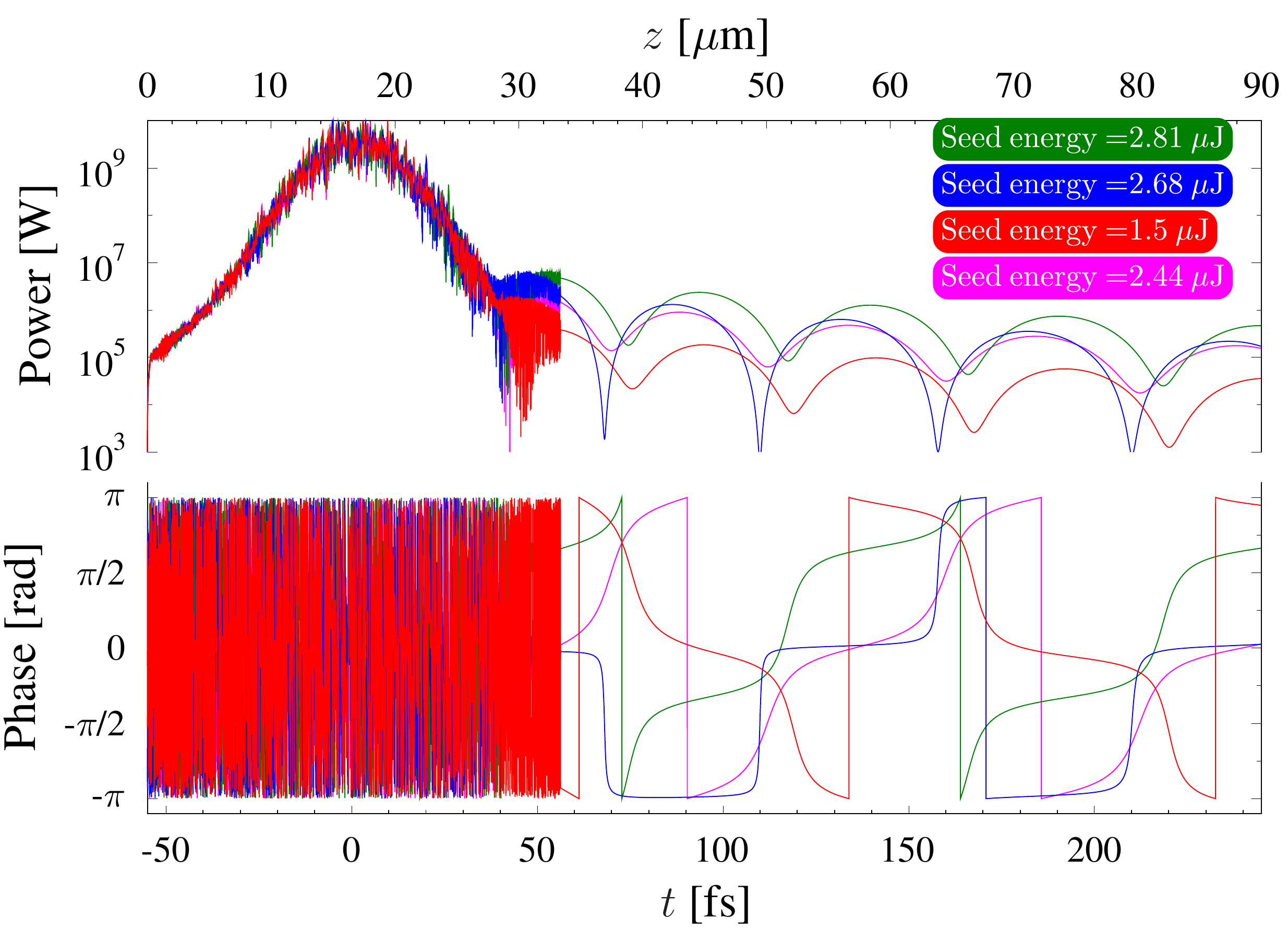}}
\end{picture}
\caption{(Color online) Time dependencies of power and phase of the entire radiation
  interacting with the electron bunch at the beginning of the U2 undulator
  system. The entire radiation comprises both the SASE radiation 
  (the first term of Eq.~\eqref{pro076}), and the monochromatic seed
  generated by FBD of the SASE radiation from a crystal (the second term
  of Eq.~\eqref{pro076}). Examples of different SASE pulses in the FEL
  long-pulse operation mode are shown, demonstrating jitter in the
  seed power, and positions of the seed maxima. The crystal
  monochromator with $\timexo=0.75$~fs is used.}
\label{Fig007}
\end{figure}

There are typical shot-to-shot SASE radiation fluctuations after the
U1 undulator system. As a result, the power and phase of the radiation
after the monochromator fluctuates as well, as shown in
Fig.~\ref{Fig007}. See also the discussion in \cite{LS12}. The power
fluctuations could be about 50\%. There is a $\simeq 10-15$-fs jitter
in the duration and in the form of the SASE signal, resulting in a
similar jitter of the seed oscillation peak positions and of the seed
phase.  The phase time dependences appear to be very different from
shot to shot, which is related to how the convolution of a particular
SASE radiation pulse with the crystal response function works, see
Eq.~\eqref{pro076}.  For realistic FEL output, averaging is required
over many shorts. The averaged picture, however, may obscure the
underlying physics. For this reason, the following analysis is
performed without averaging, but using instead a single typical SASE
radiation pulse for all cases. In particular, the SASE pulse, which
produces the seed with $2.44~\mu$J energy in Fig.~\ref{Fig007} is used
in the long-bunch simulations.

The FEL calculations are performed within a $6\bunchlengthrms$ time
window, where $\bunchlengthrms $ is the rms duration of the electron
bunch.  The current and energy distributions along the electron bunch
are assumed to be Gaussian.  Calculations in the U2 FEL system are
performed using a fresh bunch, so as not to obscure the self-seeding
effects by additional energy spread after the U1 FEL system.  Using a
fresh bunch in the U2 FEL system is actually a very attractive option
from many points of view.  Use of the fresh bunch has been discussed
in the two-electron-bunch self-seeding scheme \cite{DZR10}.

\begin{figure}[t!]
\setlength{\unitlength}{\textwidth}
\begin{picture}(1,0.9)(0,0)
\put(0.0,0.00){\includegraphics[width=0.5\textwidth]{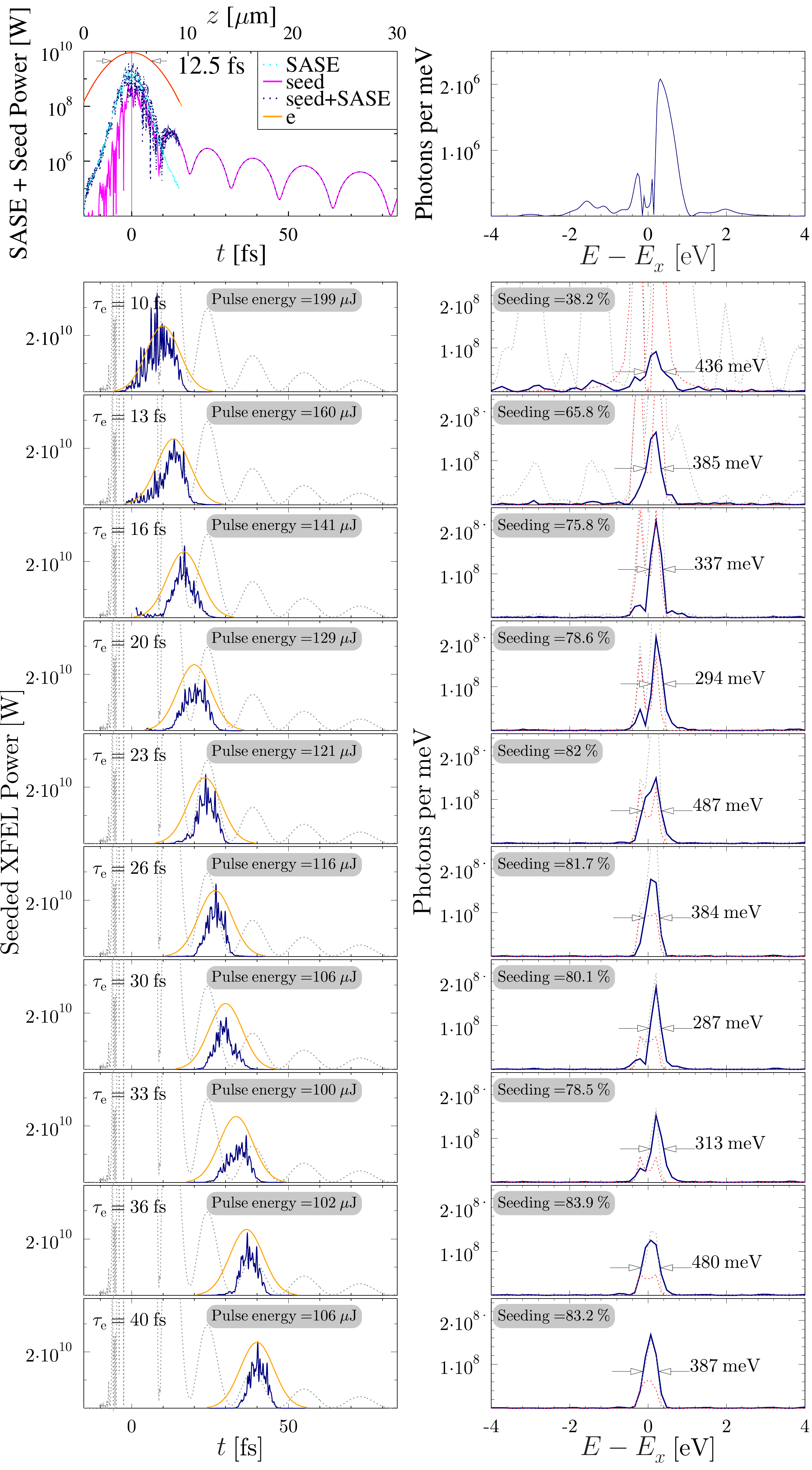}}
\end{picture}
\caption{(Color online) Time (left) and spectral (right) dependencies of the XFEL
  radiation in the self-seeding short-bunch mode. Graphs in the top
  row, present the dependencies at the entrance of the U2 undulator
  system.  Seeding is performed by the FBD monochromator with the FBD
  time constant $\timexo=0.21$~fs. The top right spectrum of the
  seeding radiation is time-integrated.  The lower graphs show in
  solid dark blue lines the time and spectral dependencies at the end
  of the U2 system for different electron bunch delays
  $\delaye$. Orange lines show the electron current profiles. The
  dashed gray lines in the left graphs are the time profile of the
  seed. The dashed gray lines in the right graphs represent the seed
  spectrum as it is ``seen'' by the electron bunch at appropriate time
  delays. Red dashed lines show the same, however, for an idealistic
  case of instantaneous excitation of the crystal with a white
  spectrum. The nominal photon energy $E_{\ind{x}}=8.3$~keV. }
\label{Fig008}
\end{figure}

\begin{figure}[t!]
\setlength{\unitlength}{\textwidth}
\begin{picture}(1,1.07)(0,0)
\put(0.0,0.00){\includegraphics[width=0.5\textwidth]{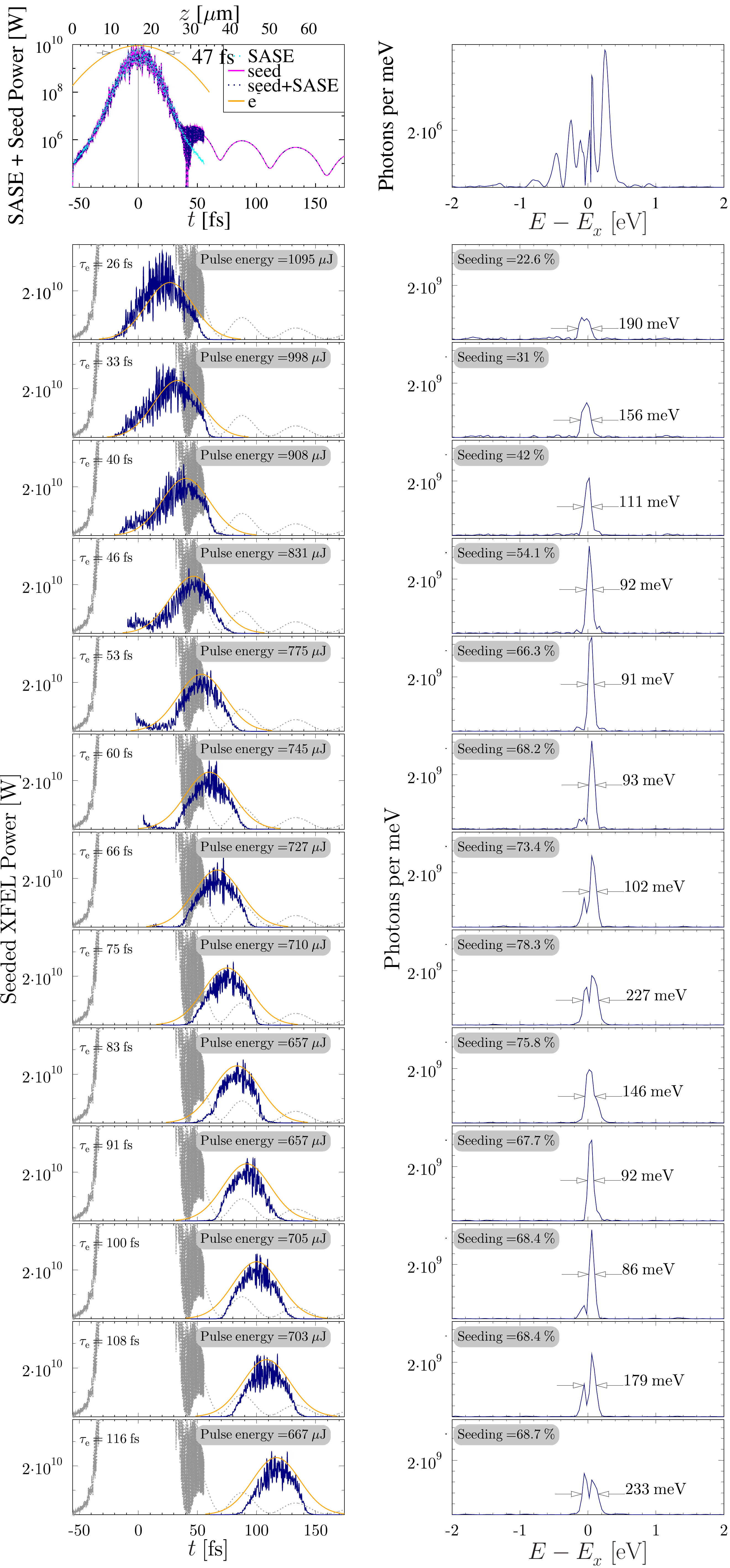}}
\end{picture}
\caption{(Color online) Similar to Fig.~\ref{Fig008}, however, showing results of
  calculations in the long-bunch mode, with seeding by the FBD
  monochromator with the FBD time constant $\timexo=0.75$~fs. 
The nominal photon energy $E_{\ind{x}}=9.1$~keV.}
\label{Fig009}
\end{figure}

The orange lines in Figs.~\ref{Fig008}, \ref{Fig009}, \ref{Fig010}, and
\ref{Fig011} show the electron current profile within this time
range. The dashed cyan lines represent the SASE radiation output after
U1 undulator system within the same time window. The solid magenta
lines represent the pure monochromatic seed - the intensity of the
second term in Eq.~\eqref{pro076}, calculated for FBD with different
$\timexo$ values. The dashed blue lines represent the entire
radiation, the sum of the SASE and of the monochromatic seed, i.e.,
everything that enters the U2 undulator system. The phase of the seed
is shown in the graphs in the $\beta$-rows of Figs.~\ref{Fig010} and
\ref{Fig011}.


\begin{table}[t!]
\centering
\begin{tabular}{|l|lll|}
  \hline  \hline  
Parameter \&  Symbol  &   Long-bunch    & Short-bunch    &  Unit \\        
  &   mode    & mode   &   \\        
  \hline  \hline  
Electron energy  $E_{\indrm{e}}$   &  14.3  &    13.6  &    GeV    \\                                                             
Bunch charge        &  150   &    40    &    pC     \\
Peak current   $I_{\indrm{e}}$     &  3     &    3     &    kA     \\
Slice energy spread &  2.8   &    5.6   &    MeV    \\
Slice emittance     &  0.4   &    0.4   &    $\mu$m \\
Bunch duration $\bunchlengthrms$  & 20 {\tiny (rms)} &   5.3 {\tiny (rms)} &  fs  \\
Bunch duration  $\bunchduration$  & 47 {\tiny (FWHM)} &   12.5 {\tiny (FWHM)} &  fs \\
U1 length           &    60 & 60 &  m \\
U2 length           &    70 & 70 &  m \\
Undulator period $\lambda_{\ind{w}}$ & 30 & 30 & mm \\ 
Normalized undulator &  &  & \\[-2mm]
parameter $a_{\ind{w}}$  & 2.4729 & 2.4729 & \\
 \hline  \hline  
\end{tabular}
\caption{Main XFEL parameters in  self-seeding mode with  two magnetic undulator systems U1 and U2, and with a FBD monochromator.}
\label{tab1}
\end{table}

The seeded power at the end of the U2 FEL system is very sensitive to
the normalized undulator parameter $a_{\ind{w}}= (e/mc) (B_{\ind{w}}/2
k_{\ind{w}})$. Variation by 0.06\% from the optimal value
$a_{\ind{w}}=2.4729$ results in a more than 50\% drop of  power in
the long-bunch mode. Parameter $a_{\ind{w}}$ is defined by the on-axis
undulator field amplitude $B_{\ind{w}}$, the undulator wave number
$k_{\ind{w}} = 2\pi/\lambda_{\ind{w}}$, the electron rest mass $m$,
and the elementary charge $e$.

\subsection{Short-bunch  mode}

We first validate the method by using the short electron bunch.  The
wake-field effects from the undulator chamber are ignored. The main
XFEL parameters are shown in Table~\ref{tab1}.  

Figure~\ref{Fig008} shows an example of calculations with a monochromatic
seed produced by filtering the SASE radiation through the FBD
monochromator with the FBD time constant $\timexo=0.21$~fs. Such a time
constant would correspond, in particular, to the 220 Bragg reflection
of the 8.3-keV x-rays from a $\thick=75-\mu$m-thick diamond crystal in
the symmetric Bragg-case geometry with an asymmetry angle of $\eta=0$, a
Bragg angle of $\theta=36.3^{\circ}$, and an extinction length of
$\extlengs=1.98~\mu$m. See Table.~\ref{tab0}.

The top left graph of Fig.~\ref{Fig008}, shows electron bunch current
profile (orange line), SASE power (dashed cyan line), seed power
(magenta solid line), and the total radiation power (dashed dark blue
line). The right top graph shows the time-integrated spectral density
of the seed after the monochromator.  It is a product of the FBD
spectrum, the kind of spectrum shown in Fig.~\ref{Fig002}(b), with a
noisy SASE spectrum.

The graphs in the lower rows show time dependencies of the XFEL
radiation power (left) and spectral densities of the XFEL radiation
(right), calculated for different electron bunch delays $\delaye$. The
spectral dependencies are shown in a range much narrower than the SASE
spectrum width, to emphasize details of the seeded spectrum.  The
bandwidths of the seeded spectra are narrower than the spectrum of the
seed, and change with the delay $\delaye$. We will present and
discuss the results of the simulations in detail in Sec.~\ref{results}.

\subsection{Long-bunch mode}

Having validated the simulations in the short-bunch mode, we apply the
procedure to the long-bunch mode. The current and energy distributions
along the bunch are also Gaussian.  For simplicity, the wake-field
effects due to the resistive wall impedance of the undulator chamber
are not considered here as well.

In the short-bunch mode, the wake-field effects are negligible.
However, in the long-bunch mode, they may induce a substantial energy
chirp along the electron bunch, move electrons off the FEL interaction
bandwidth, and thus degrade the seeding efficiency.  There are several
techniques, such as electron bunch pre-chirp \cite{CDP06}, specially
shaped vacuum chamber \cite{BS12}, etc., that could be used to
mitigate the wake-field effects.

Figure~\ref{Fig009} shows an example of calculations with a monochromatic
seed produced by filtering the SASE radiation through the FBD
monochromator with the FBD time constant $\timexo=0.75$~fs. Such time
constant could be realized in different ways, one of them presented in
Table.~\ref{tab0}: by using the 004 Bragg diffraction of the 9.1~keV
x-rays from a $\thick=100~\mu$m thick diamond crystal in the symmetric
Bragg-case geometry with an asymmetry angle $\eta=0$, a Bragg angle of
$\theta=56.8^{\circ}$, and an extinction length of
$\extlengs=3.63~\mu$m.

One of striking features, of the data presented in Fig.~\ref{Fig009},
is that the substantial variation of the seeded radiation bandwidth
with the electron bunch delay $\delaye$. We will present and discuss
the results of the simulations in detail in Sec.~\ref{results}.

\begin{table}[t!]
\centering
\begin{tabular}{|l|lll|}
  \hline  \hline  
Parameter \&  Symbol  &   Long-bunch    & Short-bunch    &  Unit \\        
  &   mode    & mode   &   \\        
  \hline  \hline  
Photon energy $E_{\indrm{x}}$ &  9.10  &    8.30  &    keV \\
Wavelength    &  1.36  &    1.50  &    $\AA$  \\
 \hline  
\multicolumn{4}{|c|}{SASE radiation after U1} \\
 \hline  
SASE pulse energy  &    83      &   9   	& $\mu$J \\
SASE peak power    &    3     &  1.2	& GW \\
 \hline  
\multicolumn{4}{|c|}{Seeded radiation after U2} \\
 \hline  
Pulse peak power          &  20  &    15  &    GW  \\
Photons per pulse   &  $4\times 10^{11}$  & $8\times 10^{10}$   &    \\
Pulse energy   &  680  & 100   &  $\mu$J  \\
Pulse duration $\xduration$     &  34 {\tiny (FWHM)}  &  8 {\tiny (FWHM)}  & fs   \\
Pulse duration $\xdurationrms$     &  15 {\tiny (rms)}  &  3.5 {\tiny (rms)}  & fs   \\
Min. bandwidth $\xband$     &  90 {\tiny (FWHM)} &  320 {\tiny (FWHM)}  & meV   \\
Min. bandwidth $\xbandrms$     &  38 {\tiny (rms)} &  136 {\tiny (rms)}  & meV   \\
Max. spectral flux $\spflux_{\indrm{m}}$ &  $\simeq 3\times10^9$     & $\simeq 2\times10^8$   & ph/meV \\  
$\spflux_{\indrm{m}} $ per pC, $\spfluxnorm_{\indrm{m}}$  &  $\simeq 2\times10^7$     &  $\simeq 6\times10^6$  & ph/meV/pC \\  
  \hline  \hline  
\end{tabular}
\caption{$\timexo$-independent XFEL radiation parameters in  self-seeding mode with FBD monochromators. 
  By ``seeded radiation'' we define the portion of the total radiation enclosed within the narrow bandwidth, as a result of self-seeding process.}
\label{tab2}
\end{table}

\section{Results and Discussion}
\label{results}

\begin{figure*}[t!]
\setlength{\unitlength}{\textwidth}
\begin{picture}(1,1.05)(0,0)
  \put(0.0,0.00){\includegraphics[width=1.0\textwidth]{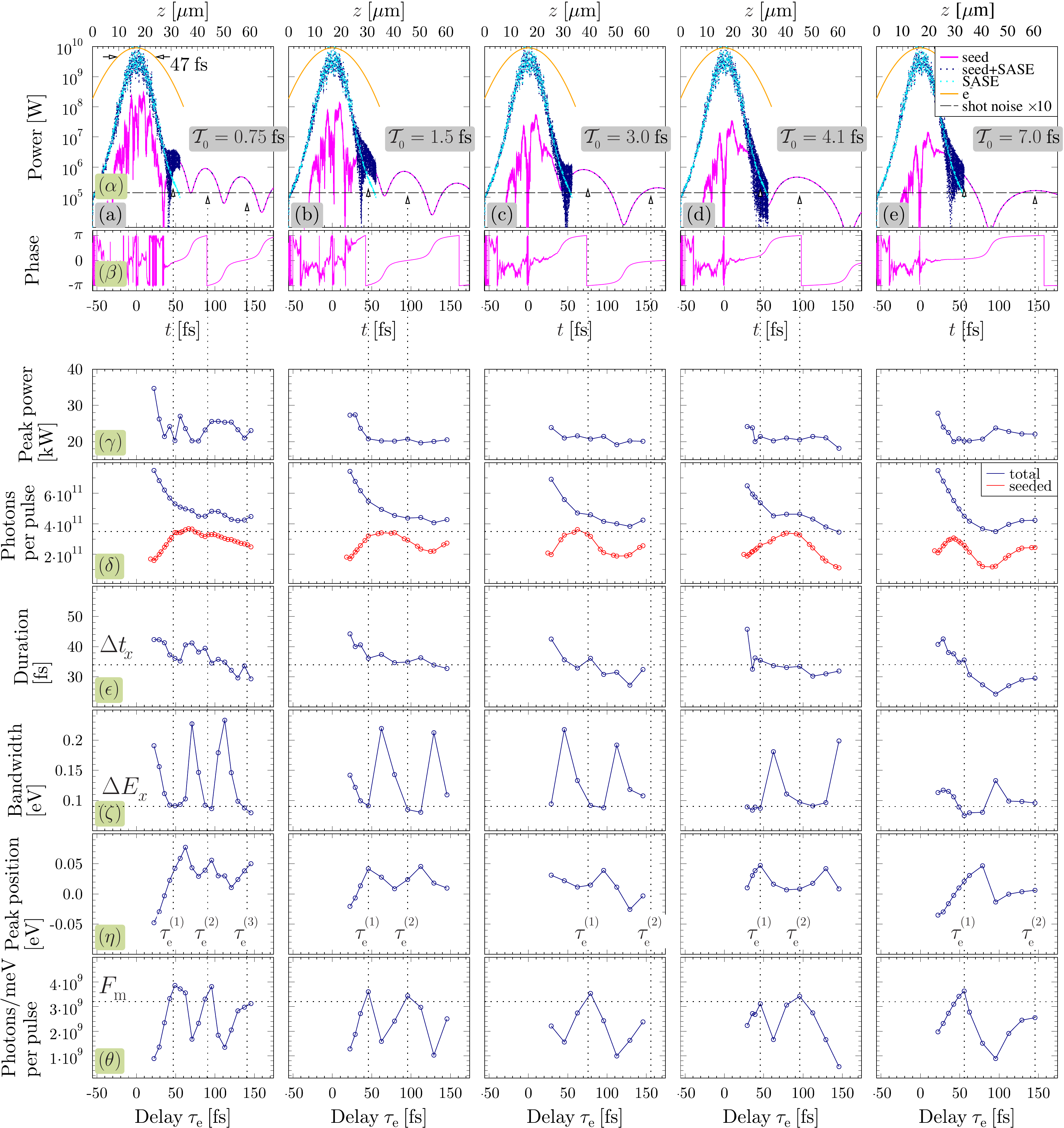}}
\end{picture}
\caption{(Color online) Time dependencies of the monochromatic seed power (magenta
  lines in the $\alpha$-row graphs), seed plus SASE power (dashed blue
  line in same graphs), ten times increased shot-noise level (dashed
  black line), and the time dependencies of the seed phase -- graphs
  in the $\beta$-row.  The seed is generated in the long-bunch mode by a
  SASE radiation from the U1 undulator system (cyan lines in the $\alpha$-row
  graphs) in forward Bragg diffraction (FBD) from crystals with FBD
  time constant $\timexo=0.75$~fs (a), $\timexo=1.5$~fs (b),
  $\timexo=3.0$~fs (c), $\timexo=4.1$~fs (d), $\timexo=8$~fs
  (e). Graphs in other rows represent properties of the radiation at
  the end of the undulator system U2, as a function of the electron bunch
  delay $\delaye $ in the magnetic chicane.}
\label{Fig010}
\end{figure*}

\begin{figure*}[t!]
\setlength{\unitlength}{\textwidth}
\begin{picture}(1,1.05)(0,0)
  \put(0.0,0.00){\includegraphics[width=1.0\textwidth]{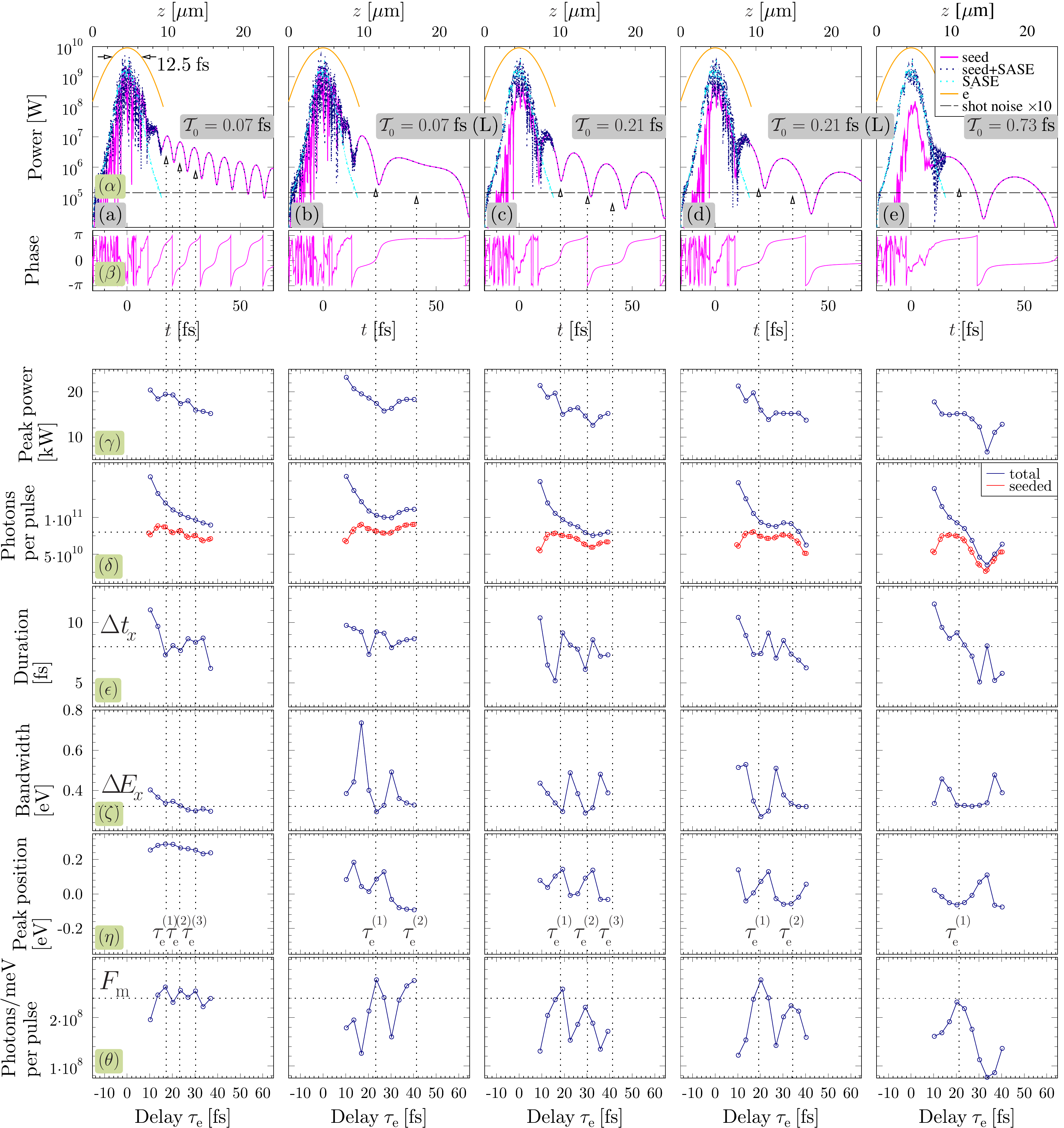}}
\end{picture}
\caption{(Color online) Same as Fig.~\ref{Fig010}, but for the short-bunch mode calculations with  (a)
  $\timexo=0.07$~fs (Bragg-case), (b) $\timexo=0.07$~fs (Laue-case),
  (c) $\timexo=0.21$~fs (Bragg-case), (d) $\timexo=0.21$~fs
  (Laue-case), (e) $\timexo=0.73$~fs (Bragg-case).
}
\label{Fig011}
\end{figure*}

The seeded XFEL radiation characteristics for different seeding
conditions are presented graphically in the long-bunch mode in
Fig.~\ref{Fig010} and in the short-bunch mode in Fig.~\ref{Fig011}.
The seeding conditions are varied by changing FBD time constant
$\timexo$ and the electron bunch delay $\delaye$. The results for
different $\timexo$ are shown in the appropriate columns of
Figs.~\ref{Fig010} and \ref{Fig011}.  The XFEL radiation
characteristics (the focus is on spectral properties) such as peak
power, total number of photons, total number of seeded photons, x-ray
pulse duration, spectral bandwidth, and spectral line center position
are presented as a function of the electron bunch delay $\delaye$ by
the graphs in respective rows of Figs.~\ref{Fig010} and
\ref{Fig011}. $\timexo$-independent characteristics of the XFEL
radiation are presented in Table~\ref{tab2}.

The main observations, which can be derived from the results presented
in Figs.~\ref{Fig010} and \ref{Fig011}, are summarized below.

\subsection{Peak Power}

The peak power of the radiation pulse at the end of the XFEL is
stable, and independent of $\timexo$ and of the electron bunch delay
$\delaye$. See graphs in the $\gamma$-rows. It has approximately the same
magnitude, both in the short-bunch and in the long-bunch modes. Still,
it is slightly larger in the long-bunch mode. This is consistent with
the FEL theory, see Eq.~(91) of \cite{HK07}, predicting the saturation
power level $P_{\indrm{sat}}\propto \rho P_{\indrm{beam}}$ to be a
fraction of the total electron beam power
$P_{\indrm{beam}}=E_{\indrm{e}} I_{\indrm{e}}/e$, with the fraction
given by the dimensionless Pierce parameter $\rho$ \cite{BPN84}, which
is typically $\simeq 10^{-3}$.  The peak current $I_{\indrm{e}}$ and
$\rho$ values are the same in both cases, however, the electron energy
$E_{\indrm{e}}$ is chosen larger in the long-bunch mode, see
Table~\ref{tab1}.

\subsection{Number of Photons per Pulse}

The total number of photons per pulse changes with the delay
$\delaye$. It is the largest at smaller delays, where the SASE
contribution is the largest, see graphs in the $\delta$-rows of
Figs.~\ref{Fig010} and ~\ref{Fig011}. On average, the total number of
photons is four times larger in the long-bunch mode, which corresponds
to a four times larger pulse duration and four times larger charge in
the long-bunch compared to the short-bunch mode.

We are interested primarily in the total number of ``seeded photons'',
those condensed in a narrow bandwidth, as a result of the self-seeding
process. The bandwidth may change with $\timexo$, as seen in the right
columns of Figs.~\ref{Fig008} and ~\ref{Fig009}. The total number of
seeded photons within a band defined as the four smallest bandwidths
is shown by red circles in $\delta$-row-graphs of Figs~\ref{Fig010},
and \ref{Fig011}. The total number of seeded photons is largest for
those electron bunch delays $\delaye$ at which the electron bunch is
seeded with the radiation (magenta lines in the $\alpha$-row graphs of
Figs.~\ref{Fig010} and \ref{Fig011}) whose power is much larger than
the shot-noise power $P_{\indrm{n}}$ level. Dashed black lines in the
$\alpha$-row graphs of Figs.~\ref{Fig010} and \ref{Fig011} show the
$10\times P_{\indrm{n}}$ level.  The effective shot-noise power is
estimated to be $P_{\indrm{n}} \simeq \rho^2
E_{\indrm{e}}E_{\indrm{x}}/\hbar\sqrt{2\pi}\simeq 14$~kW
\cite{Kim86,HK07}.

\subsection{Pulse Duration}

The duration of the x-ray pulse $\xduration$ (FWHM) is shorter than
the duration of the electron bunch; see Figs.~\ref{Fig008} and
\ref{Fig009}, and the graphs in the $\epsilon$-rows of Figs.~\ref{Fig010} and
\ref{Fig011}.  This fact reflects the non-linear nature of lasing
physics, in which the high-current portion of the electron bunch
produces the most significant effect.

\subsection{Spectral Bandwidth}

The spectral bandwidth of the radiation achieves its smallest values
$\xband$ (FWHM) at specific
electron bunch delays $\delaye^{\ind{(n)}}$, with respect to the
monochromatic seed, indicated by vertical dashed lines with arrows;
see the $\zeta$-row graphs in Figs.~\ref{Fig010} and \ref{Fig011}.  The
smallest achievable bandwidth $\xband$ is practically independent of
$\timexo$. However, it takes different values in the long-bunch and in
the short-bunch modes.

The product $\xdurationrms \xbandrms\approx 5 \times
10^{-16}$~eV$\cdot$s of the rms x-ray pulse duration, and the rms
minimal x-ray pulse bandwidth, is very close to Planck's constant
$\hbar=6.6 \times 10^{-16}$~eV$\cdot$s. This suggests that
self-seeding results in the almost Fourier-transform limited, i.e., in
the fully longitudinally coherent, FEL radiation.  This also very
clearly suggests that the ultimate spectral properties of the
self-seeding XFEL are determined first of all by the length
$\bunchlengthrms $ of the electron bunch rather than by the properties
of the radiation seed. The seed helps to achieve the best performance;
however, it does not determine the final result.

The bandwidth may increase by a factor of three from its smallest
value if the electron bunch delay is chosen somewhere between the
specific delays $\delaye^{\ind{(n)}}$; see
Fig.~\ref{Fig010}($\eta$). The larger $\timexo$ is, the broader are
the $\delaye$ intervals where the smallest bandwidth values are
attained. Because of the jitter in the SASE radiation, resulting in a
jitter of the positions and of the seed power peaks and phase (see
Fig.~\ref{Fig007}), it may be more advantageous to choose FBD with
longer $\timexo$ values, to achieve stable self-seeding with narrowest
bandwidth and highest spectral flux, independent of SASE jitter. At
the same time adequate seed power should be ensured as well.  FBD with
too long a $\timexo$ may produce insufficient seed power. The optimal
value is $\timexo \simeq \bunchlengthrms/\pi$, as was discussed in
Sec.~\ref{optimal}. FBD with a very short $\timexo \ll
\bunchlengthrms/\pi$ may result in double-color seeding.

\subsection{Double-Color Seeding}

\begin{figure}[t!]
\setlength{\unitlength}{\textwidth}
\begin{picture}(1,0.25)(0,0)
\put(0.0,0.00){\includegraphics[width=0.5\textwidth]{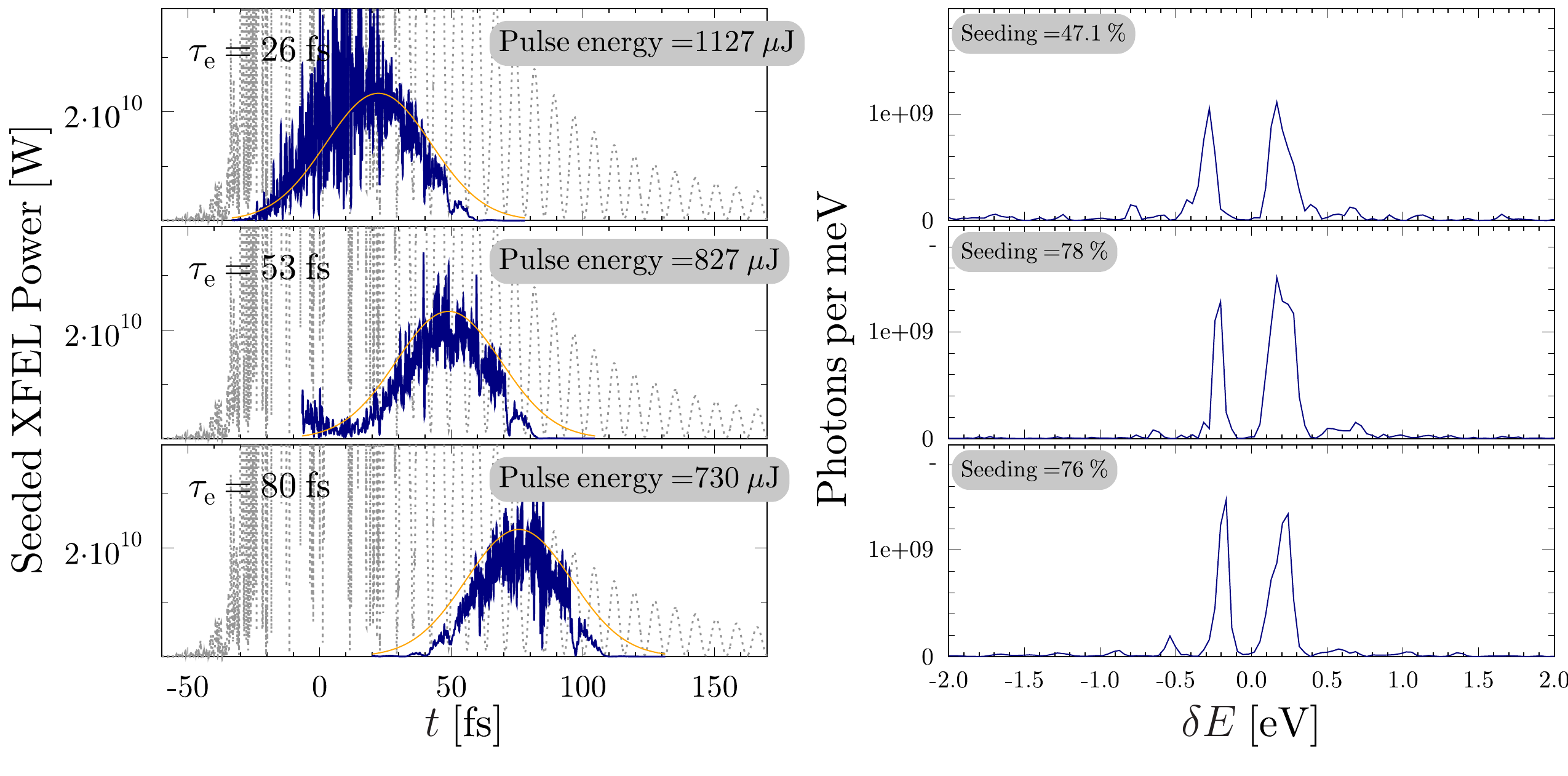}}
\end{picture}
\caption{(Color online) Time (left) and spectral (right) dependencies of the XFEL
  radiation in the self-seeding long-bunch mode with the FBD
  monochromator having $\timexo=0.07$~fs.  The graphs show in solid
  dark blue lines the time and spectral dependencies at the end of the
  U2 system for different electron bunch delays $\delaye$. Orange
  lines show the electron current profiles. The dashed gray lines in
  the left graphs are the time profiles of the seed. 
The nominal photon
  energy $E_{\ind{x}}=9.1$~keV. }
\label{Fig012}
\end{figure}

Double-color seeding in the long-bunch mode is shown in
Fig.~\ref{Fig012}, with an FBD monochromator having
$\timexo=0.07$~fs. We note that the double-color seeding with equal
intensities of both colors, as shown in Fig.~\ref{Fig012}, cannot be
achieved in each XFEL shot. Our simulations show that the relative
intensity variations of the two colors take place from shot to
shot. The success for equal-intensity double-color seeding depends on
the spectral composition of the SASE spectrum.  Only if both sharp
peaks of the FBD spectrum, as in Fig.~\ref{Fig006}(b), are
``populated'', equal-intensity double-color seeding does take
place. The total power of the seeded radiation is the same as in
single-color seeding; however, it is shared between the two spectral
components.

\subsection{Spectral Line Center Position}
\label{spectrallineposition}

The center of the self-seeding XFEL output spectrum varies with the
electron bunch delay, as shown in the $\eta$-row graphs of Figs.~\ref{Fig010}
and \ref{Fig011}.  The variation is not pronounced; however, it
correlates with the time structure of the monochromatic seed.

A considerable shift of $\simeq +0.2$~eV of the peak position in the
$\timexo=0.07$-fs Bragg case in the short-bunch mode is due to the
fact that this is actually the double-color seeding case.  There are
actually two sharp peaks in this case FBD seeding spectrum, as in
Fig.~\ref{Fig006}. However, for the particular SASE spectrum used in
theses simulations, only one of the two peaks is ``populated''.

\subsection{Spectral Flux}

In each mode, the maximum spectral flux $\spflux_{\indrm{m}}$
(measured in photons/pulse/meV) is independent of $\timexo $; see the
$\eta$-row graphs in Figs.~\ref{Fig010} and \ref{Fig011}. In the
long-bunch mode $\spflux_{\indrm{m}}\simeq 3.2\times
10^{9}$~ph/meV/pulse, while in the short-bunch mode it is an order of
magnitude smaller -- $\spflux_{\indrm{m}}\simeq 2.2\times
10^{8}$~ph/meV/pulse.  The spectral flux values, normalized to the
total bunch charge, are $\spfluxnorm_{\indrm{m}}\simeq 2.1\times
10^{7}$~ph/meV/pC/pulse in the long-bunch mode and
$\spfluxnorm_{\indrm{m}}\simeq 0.55\times 10^{7}$~ph/meV/pC/pulse in
the short-bunch mode, respectively, and differ by a factor of
four. For comparison, an XFEL in the oscillator configuration produces
$\spfluxnorm_{\indrm{m}}\simeq 2\times 10^{7}$~ph/meV/pC/pulse
\cite{LSKF11}, the number close to the high-gain XFEL in the
long-bunch mode considered here.

The maximum spectral flux is attained at the same specific electron
bunch delays $\delaye^{\ind{(n)}}$ at which the smallest spectral
bandwidth is achieved. These delays correlate but do not always
coincide with the seed power peaks in the $\alpha$-row graphs. The
delays $\delaye$ are systematically shifted to later times from the
peaks' positions.  This is clearly seen especially in the short-pulse
mode examples in Fig.~\ref{Fig011}.

The efficient seeding can be performed by using the first or second
trailing peaks of the FBD radiation, as discussed in this paper and in
the literature \cite{GKS11,HXRSS12}.  However, it also can be
performed efficiently by using the zeroth maximum of the FBD
radiation, as the case of $\timexo=7$~fs in the long-bunch mode in
Fig.~\ref{Fig010}($\eta$) demonstrates, with the delay set at $\delaye
\simeq 50$~fs.

Figure~\ref{Fig011}($\eta$), for $\timexo=0.07$~fs, shows an
exceptional case of seeding, resulting in high spectral flux and
smallest bandwidth practically {\em independent} of the delay
$\delaye$. This does not fit into the general picture discussed until
now.  Indeed, this is achieved under exceptional conditions, when the
period of the FBD power oscillations is much shorter than the electron
bunch duration, i.e., under the conditions where double-color seeding
is expected. As was already mentioned in
Sec.~\ref{spectrallineposition}, there are actually two sharp peaks in
the FBD seeding spectrum in this case, as in
Fig.~\ref{Fig006}. However, the particular SASE spectrum used in the
simulations has spectral components which ``populate'' only one of the
of the two FBD peaks.

The case $\timexo=0.07$~fs proves again that a strong seed will
produce a stable $\delaye$-independent output with the spectral flux
and spectral width defined eventually by the XFEL parameters, rather
than by the seed properties.

In Sec.~\ref{seed-calculations} we have stated that for time delays $t
\ll \thicknessduration$, the Bragg-case and Laue-case FBDs are
equivalent and are defined by a single parameter -- the FBD time
constant $\timexo$; see Eq.~\eqref{rta1250}.  At longer times,
however, Bragg-case and Laue-case FBDs are not equivalent, and have to
be described by different equations~\eqref{rta125} and \eqref{lta050},
respectively, with the time constant $\thicknessduration$ taken into
account.  Examples of the $\timexo=0.07$~fs Bragg-case and the
$\timexo=0.07$~fs Laue-case, in the short-bunch mode, with
$\thicknessduration=94$~fs (see Table~\ref{tab0}), demonstrate
substantial differences between the two cases already for time delays
$t>30$~fs. Bragg-case FBD results in double-color seeding.  On the
contrary, Laue-case FBD ensures single-color seeding with very good,
if not the best, results for spectral density and spectral width, due
to a strong signal and large duration of the seed at $t=40$~fs.

\section{Applications}

Self-seeding XFELs can generate fully coherent x-ray pulses with a
very high spectral flux, much higher than the state-of-the-art storage
ring-based synchrotron radiation sources.  Their supreme spectral
properties will expand the science reach of the XFELs and stimulate
the utilization of advanced high-resolution spectroscopic techniques.

A self-seeding XFEL, in the long-bunch mode considered in this paper,
can generate in each pulse $\simeq 2.4\times 10^{11}$~photons
concentrated within an $80$-meV broad (FWHM) line, or $\simeq 2.9\times
10^{9}$~photons/meV. With the 120-Hz repetition rate of the LCLS
machine, the average number of photons that can be generated in an
$80$-meV broad (FWHM) line is $\simeq 2.8\times 10^{13}$~photons/s,
corresponding to a spectral flux of $\simeq 3.5\times
10^{11}$~photons/s/meV.  This spectral flux number is an order of
magnitude larger than an average spectral flux available at the 3rd
generation storage ring-based machines such as ESRF, APS, or
SPring-8. The future high-repetition-rate XFELs, such as the European XFEL
\cite{Altarelli:1088597}, with an average 30-kHz pulse rate, will be
able to generate a spectral flux of $ \simeq 8.7\times
10^{13}$~photons/s/meV, i.e., approximately three orders of magnitude
more than the average spectral flux available presently at the
synchrotron radiation facilities.

High-resolution inelastic x-ray scattering (IXS) and resonant IXS
(RIXS) spectroscopies, widely used for studies of collective
excitation in condensed matter \cite{Burkel2,KS07,AVD11}, will
tremendously benefit from x-ray sources with largely enhanced average
and peak spectral flux. In particular, applications of IXS and RIXS
spectrographs working with 9-keV photons \cite{Shvydko12,SSM13} will
permit studies with spectral resolution better than 0.1~meV. X-ray
sources with largely enhanced average spectral flux would be also
attractive for other high-spectral-resolution tools, such as nuclear
resonance (M\"ossbauer) spectroscopies \cite{GdW,RR04}, as was
discussed, in particular, in the first paper on hard x-ray
self-seeding \cite{SSSY}.

\section{Conclusions}

We have performed theoretical studies for the best conditions under
which stable self-seeding of the hard x-ray FEL can be achieved with
smallest shot-to-shot variations of the output radiation, with
narrowest bandwidth, and largest spectral flux.  Characteristic time
$\timexo$ of forward Bragg diffraction (FBD) in the x-ray
monochromator crystal determines the power, spectral, and time
characteristics of the FBD monochromatic seed.  For a given electron
bunch duration $\bunchlengthrms $ the spectral flux of the
self-seeding XFEL can be maximized, the spectral bandwidth can be
respectively minimized, by choosing $\timexo\sim \bunchlengthrms/\pi
$, and by optimizing the electron bunch delay $\delaye$.  The choice
of $\timexo$ and $\delaye$ is not unique. In all cases, the maximum
value of the spectral flux and the minimum bandwidth are primarily
determined by the electron bunch length $\bunchlengthrms $. The longer
the electron bunch, the higher spectral flux can be achieved. It is
important, however, that the power of the seed is about two orders of
magnitude larger than the shot noise power, which calls for using FBD
monochromators with smaller $\timexo$ values. However, if
$\timexo\ll\bunchlengthrms/\pi $ is very short, this may result in 
two-color seeding.

\section{Acknowledgments}

We gratefully acknowledge useful discussions with ZhiRong Huang,
Gregory Penn, L.H. Yu, and Sven Reiche.  ZhiRong Huang and Juhao Wu
are acknowledged for providing LCLS FEL parameters. Work at Brookhaven
National Laboratory was supported by the U.S. Department of Energy,
Office of Science, Office of Basic Energy Sciences, under Contract
No. DE-AC02-98CH1-886.  Work at Argonne National Laboratory was
supported by the U.S. Department of Energy, Office of Science, under
Contract No. DE-AC02-06CH11357.


\end{document}